\documentclass[useAMS,usegraphicx]{mn2e}

\newcommand\CI{C\,{\sc i}}
\newcommand\OI{O\,{\sc i}}
\newcommand\CII{C\,{\sc ii}}
\newcommand\HI{H\,{\sc i}}
\newcommand\NaI{Na\,{\sc i}}
\newcommand\MgI{Mg\,{\sc i}}
\newcommand\MgII{Mg\,{\sc ii}}
\newcommand\NHI{NH\,{\sc i}}
\newcommand\nHI{n$_{HI}$}
\newcommand\NHtwo{NH$_2$}
\newcommand\nHtwo{n$_{H_2}$}

\newcommand\nH{n$_H$}
\newcommand\Htwo{H$_2$}
\newcommand\mum{$\mu$m~}

\newcommand\kms{km~s$^{-1}$}

\newcommand\cmtwo{cm$^{-2}$}

\newcommand\tCO{$^{12}$CO}

\newcommand\cmthree{cm$^{-3}~$}

\newcommand\etal{et al.}
\newcommand\be{\begin{equation}}
\newcommand\ee{\end{equation}}
\newcommand\bea{\begin{eqnarray}}
\newcommand\eea{\end{eqnarray}}

\newcommand\ddeg{$^{o}$}

\catcode`\@=11
\newcommand{\gsim}{${\mathrel{\mathpalette\@versim>}}$}
\newcommand{\lsim}{${\mathrel{\mathpalette\@versim<}}$}
\newcommand{\@versim}[2]{\lower 2.9truept \vbox{\baselineskip 0pt \lineskip
    0.5truept \ialign{$\m@th#1\hfil##\hfil$\crcr#2\crcr\sim\crcr}}}
\catcode`\@=12

\begin{document}

\title[Carbon RRLs toward Cold \HI\ Regions]
{Carbon Recombination Lines toward the Riegel-Crutcher Cloud and other
Cold \HI\ Regions in the inner Galaxy \\}

\author[D. Anish Roshi \& N. G. Kantharia] 
{D. Anish Roshi, $^1$\thanks{aroshi@nrao.edu} \& N. G. Kantharia, $^2$ 
\thanks{ngk@ncra.tifr.res.in} \\
$^1$ 
National Radio Astronomy Observatory\footnote{National Radio Astronomy Observatory is a facility
of the National Science Foundation operated under cooperative agreement by Associated Universities, Inc.}, 
Green Bank, WV 24944, USA \\
Raman Research Institute, Bangalore, India 560 080 and  \\
National Centre for Radio Astrophysics, Tata Institute of Fundamental Research, 
Pune, India.  \\
$^2$ National Centre for Radio Astrophysics, Tata Institute of Fundamental Research,
Pune, India\\
}

\maketitle


%


%

\begin{abstract}

In the first paper in the series, Roshi, Kantharia \& Anantharamaiah (2002) published
the Galactic plane survey of carbon recombination lines (CRRL) at 327 MHz.  
CRRL were extensively detected from the inner Galaxy (longitudes$ < 20^{\circ}$).  
We report here, for the first time, the association of low frequency CRRL with
\HI\ self-absorbing clouds in the inner Galaxy and that
the CRRLs from the innermost $\sim 10^{\circ}$ of the Galaxy arise in the Riegel-Crutcher (R-C) cloud.
The R-C cloud is amongst the most well known of \HI\ self-absorbing (HISA) regions 
located at a distance of about 125 pc in the Galactic centre direction. 
Taking the R-C cloud as an example, we demonstrate that 
the physical properties of the HISA can be constrained 
by combining multi-frequency CRRL and \HI\ observations.  
The derived physical properties of the HISA cloud are used to determine the cooling and heating rates. The 
dominant cooling process is emission of the \CII\ 158 \mum line whereas 
dominant heating process in the cloud interior is photoelectric emission. 
Constraints on the FUV flux (G0 $\sim$ 4 to 7) falling on the R-C cloud are obtained by 
assuming thermal balance between the dominant heating and cooling processes.
The H$_2$ formation rate per unit volume in the cloud interior is 
$\sim$ 10$^{-10}$ -- 10$^{-12}$ s$^{-1}$ \cmthree,
which far exceeds the H$_2$ dissociation rate per unit volume.  We conclude that
the self-absorbing cold \HI\ gas in the R-C cloud may be in the process of converting 
to the molecular form. The cold \HI\ gas observed as HISA features are ubiquitous
in the inner Galaxy and form an important part of the ISM. Our analysis 
shows that combining CRRL and \HI\ data can give important insight into the nature
of these cold gas. We also estimate the integration times required to 
image the CRRL forming region with the upcoming SKA pathfinders. Imaging with
the MWA telescope is feasible with reasonable observing times. 
\end{abstract}

\section{Introduction}

In earlier papers (\nocite{ra00}Roshi \& Anantharamaiah 2000, \nocite{ra01a}2001a, \nocite{ra01b}2001b),
we presented the details of a 327 MHz recombination line survey of the inner Galaxy 
made with the Ooty Radio Telescope (ORT). Results of the preliminary analysis of  
carbon recombination lines (CRRL) detected in this survey were 
presented in \nocite{rka02}Roshi, Kantharia, Anantharamaiah (2002; Paper I).
The CRRLs at low-frequencies (\lsim\ 1.4 GHz) arise in diffuse \CII\ regions in the Galaxy.  
The ionisation potential of carbon (11.3 eV) is less than that
of hydrogen (13.6 eV) and hence carbon can remain in the singly-ionised state
outside regions where hydrogen is fully ionised.
Low-frequency CRRLs are thus useful as diagnostics of partially ionised clouds.
However, the reduced abundance of carbon (solar abundance relative to hydrogen $2.9 \times 10^{-4}$; 
\nocite{l03}Lodders 2003) and the consequent weak strength of the CRRLs makes the detection of these
lines difficult. Generally stimulated emission
or absorption against a strong background radiation field facilitate their detection.  

The first CRRL from diffuse \CII\ region was detected toward the direction of the
supernova remnant Cas~A and
this direction has since been extensively observed in CRRL at frequencies ranging 
from 15 MHz in absorption to 1400 MHz in emission
(\nocite{ks80}Konovalenko \& Sodin 1980, \nocite{bcw80}Blake Crutcher \& Watson,
1980, \nocite{k84}Konovalenko 1984, \nocite{k90}1990, \nocite{eetal84}Ershov \etal\ 1984, 
\nocite{eetal87}1987, \nocite{pae89}Payne, Anantharamaiah, Erickson
1989, \nocite{sw91}Sorochenko \& Walmsley 1991, \nocite{ssetal07}Stepkin et al. 2007). 
Modelling showed that CRRLs originate in cool gas with $T_e \sim 35-75$K (\nocite{pae94}Payne 
 \etal\ 1994). The LSR velocity coincidence of the
carbon lines with \HI\ absorption observed toward Cas~A and the similarity in
the spatial distribution of the two lines across Cas~A have led to the suggestion 
that the CRRL forming regions and \HI\ absorption regions coexist 
(\nocite{kap98}Kantharia \etal\ 1998). 
CRRLs from the Galactic plane region 
have also been detected near 327 MHz (Roshi \& Anantharamaiah 2000,2001a),
near 76 MHz (Erickson et al. 1995), near 34.5 MHz (Kantharia \& Anantharamaiah 2001) and
near 26 MHz (private communication: S. Stepkin).  Most of these have been confined to the
inner Galaxy at longitudes less than $20^{\circ}$ and arise in extended diffuse CII regions
coincident with \HI\ regions.  CRRLs have thus been
detected towards several directions in the Galactic plane, however the partially ionised gas 
towards Cas~A remains the best studied region. 

The sight line towards Cas~A intercepts the Orion and Perseus arms of the Galaxy 
and the \HI\ absorptions detected against Cas~A are due to the cold neutral medium (CNM)
in this direction.  Much of the atomic hydrogen in the Galaxy has been observationally found to exist 
as `warm' ($\sim 10^4$ K; warm neutral medium) and `cold' ($\sim 70$ K; CNM) gases. 
Models of the ISM indicate that these two temperature
gases coexist in pressure equilibrium (\nocite{fgh69}Field, Goldsmith \& Habing 1969) 
with a mean pressure, measured in the solar neighbourhood, of 2240 K \cmthree
(\nocite{jt01}Jenkins \& Tripp 2001; see also \nocite{wetal03}Wolfire \etal\ 2003;
\nocite{kh88}Kulkarni \& Heiles 1988, \nocite{dl90}Dickey \& Lockman 1990). The cold \HI\ is 
observationally studied using 21cm absorption lines
against background continuum sources (eg. \nocite{ht03}Heiles \& Troland 2003) 
as well as \HI\ ``self-absorption'' (HISA; eg. \nocite{k74}Knapp 1974, \nocite{getal00}Gibson \etal\ 2000, 
\nocite{ketal05}Kavars \etal\ 2005). 
HISA is due to \HI\ absorption by cold atomic gas against bright background 21cm emission.
While the temperature of cold gas is constrained by \HI\ absorption
studies (\nocite{ht03}Heiles \& Troland 2003), the gas pressure is measured using the ultraviolet 
absorption lines of CI (\nocite{jt01}Jenkins \& Tripp 2001). The cooling of the cold gas,
which is predominantly due to \CII\ (the most abundant gas phase ion in CNM) fine-structure 
transition, has been studied through observations of the 158 \mum line emission 
(eg. \nocite{betal93}Bock \etal\ 1993).
Attempts have also been made to study the cold \HI\ gas using the CRRL emission near 1.4 GHz
(\nocite{c77}Crutcher 1977).

In this paper,  we suggest, for the first time, 
the association of low frequency CRRL with
the cold \HI\ gas observed as HISA features in the Galactic plane. 
Stimulated emission due to the galactic background continuum 
radiation facilitates detection of CRRLs in emission near 327 MHz
from this cold gas. We also, for the first time, suggest that the CRRLs detected
from the Galactic centre direction arise in 
the Riegel-Crutcher (R-C) cloud (\nocite{rc72}Riegel \& Crutcher 1972), a prominent HISA. 
Although earlier studies have indicated
that the carbon lines are formed in the same LSR velocity range over which
\HI\ absorption and $^{12}$CO emission are observed in the inner Galaxy 
(\nocite{ema95}Erickson, McConnell, Anantharamaiah 1995, \nocite{ka01}Kantharia \& Anantharamaiah 2001)
the association of CRRLs  with HISA in the inner Galaxy
has not been discussed earlier. A summary of the 327 MHz recombination data is given in Section~\ref{crrlobs},
which is followed by a discussion on the association of CRRL with cool atomic gas 
(Section~\ref{crrlasso}). In Section~\ref{rccloud} we focus on the R-C cloud, 
and  demonstrate that CRRL and \HI\ data can be combined to determine 
the physical properties of the line forming region. In Section~\ref{hcm} we use the estimated
physical properties to investigate heating and cooling of the gas in the R-C cloud 
and molecule formation in the cloud. The results are summarised in Section~\ref{sum}.   

\section{Summary of our CRRL data}
\label{crrlobs}

The ORT recombination line survey data were obtained with two angular resolutions 
($2^{\circ} \times 2^{\circ}$; \nocite{ra00}Roshi \& Anantharamaiah 2000;  $2^{\circ} \times 6'$;
\nocite{ra01a}Roshi \& Anantharamaiah 2001a).
A galactic longitude range $-28^{\circ} < l < 89^{\circ}$ and 
latitude $b = 0$\ddeg\ was covered in the low resolution ($2^{\circ} \times 2^{\circ}$) survey. 
CRRLs have been  detected almost contiguously from
$-2^{\circ} < l < 20^{\circ}$ and also in a few positions at other longitudes (Paper I). 
A few of these positions were then `mapped' with the high resolution beam ($2^{\circ} \times 6'$;
\nocite{ra01a}Roshi \& Anantharamaiah 2001a). 
In Paper I, we discussed some of the results from this survey.  Summarising, we find that the
radial distribution of the CRRLs near 327 MHz is similar to that of star-forming regions 
traced by the 3 cm hydrogen RRLs (\nocite{l89}Lockman 1989) and $^{12}$CO (\nocite{detal87}Dame \etal\ 1987).
Our multi-resolution ORT data also indicate
that some of the diffuse \CII\ regions have an angular extent of a few degrees.

\section{CRRLs and \HI\ self-absorption regions}
\label{crrlasso}


\begin{figure}
\centering
\includegraphics[width=5in, height=1.97in, angle = -90]{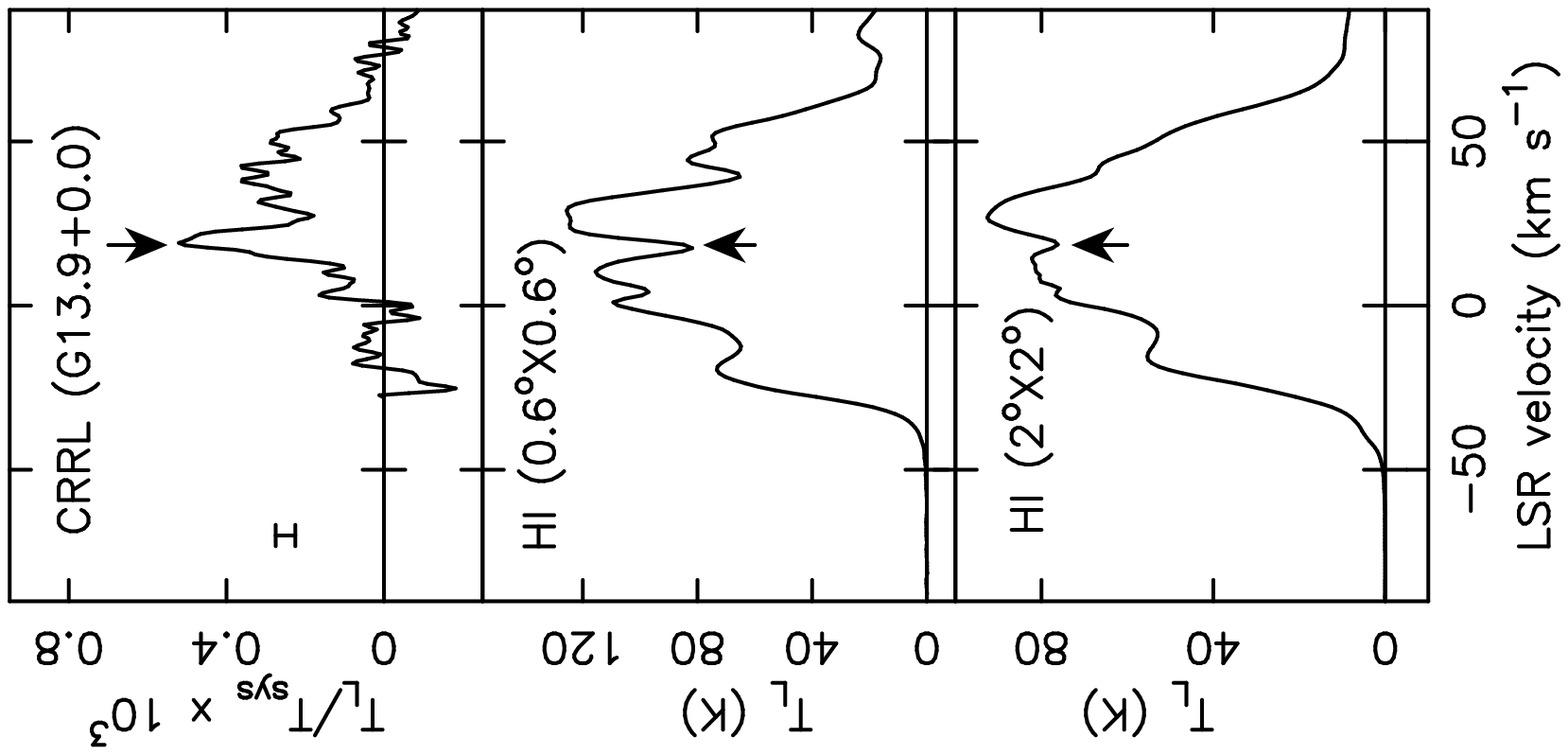}
\caption{CRRL and \HI\ spectra toward G13.9+0.0.
CRRL (top panel) spectra at 327 MHz are obtained with an 
angular resolution of 2\ddeg\ $\times$ 2\ddeg.
The 1$\sigma$ value of the spectral noise is also
shown in the top panel. 
The middle and bottom panels show \HI\ spectra with 
angular resolutions 0\ddeg.6 $\times$ 0\ddeg.6
and 2\ddeg\ $\times$ 2\ddeg\ respectively (Kalberla \etal\ 2005). 
The low resolution \HI\ spectrum is centred at the
galactic coordinates $l$ = 14\ddeg.0 and $b$ = 0\ddeg. The arrows 
are placed at the LSR velocity of the `narrow' carbon lines. 
The good LSR velocity coincidence of `narrow' CRRL and HISA 
features suggests an association between the two line forming
regions.   
}
\label{fig1}
\end{figure}

\begin{figure}
\includegraphics[width=5.0in, height=3.2in, angle = -90]{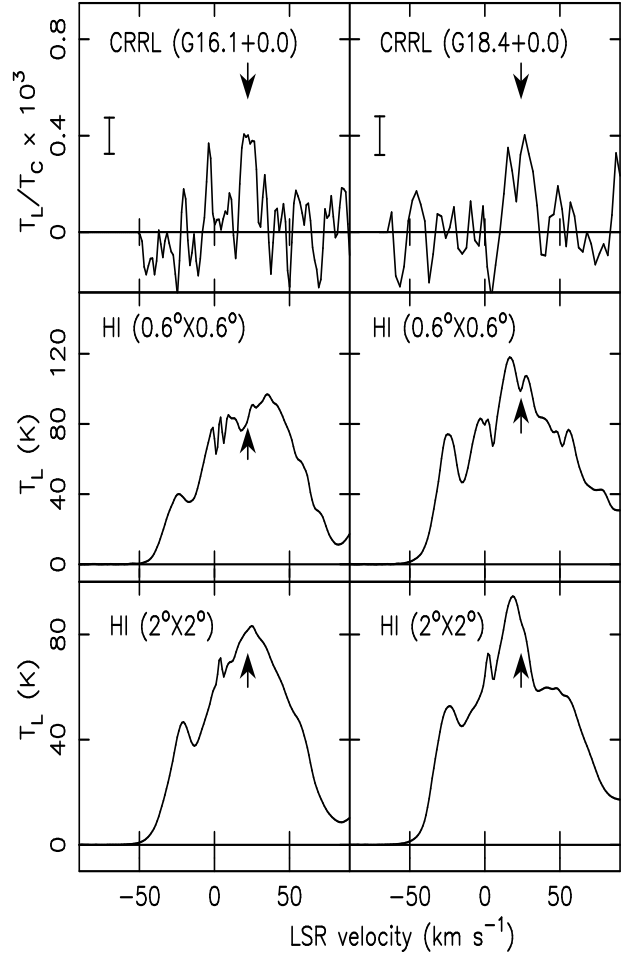}
\caption{Examples of CRRL and \HI\ spectra (Kalberla \etal\ 2005) 
toward G16.1+0.0 and G18.4+0.0 which show `narrow' CRRL line
(see top panels) emission but no HISA feature in the 2\ddeg\ $\times$ 2\ddeg\ 
averaged \HI\ spectra (bottom panels). CRRL (top panels) spectra 
at 327 MHz are obtained with an angular resolution of 2\ddeg\ $\times$ 2\ddeg.
The 1$\sigma$ values of the spectral noise are also
shown in the top panels. 
The middle panels show \HI\ spectra with angular resolutions 0\ddeg.6 $\times$ 0\ddeg.6.
The low resolution \HI\ spectra are centred at the
galactic coordinates $l$ = 16\ddeg.0 and 18\ddeg.5 and $b$ = 0\ddeg. The arrows 
are placed at the LSR velocity of the carbon lines detected at the two positions. 
}
\label{fig2}
\end{figure}

\begin{table*}
\small
\centering
\begin{minipage}{140mm}
\caption{``Narrow'' 327 MHz CRRL and \HI\ line parameters } 
\label{tab1}
\begin{tabular}{llrrlrrc}
\hline
       & \multicolumn{3}{c}{CRRL$\:^{1}$}              & \multicolumn{3}{c}{\HI\ $\:^{1}$}  &       \\
Source & $T_L/T_{C}\:^{2}$ & $\Delta V$ &  $V_{LSR}$  & $T_L$ & $\Delta V$ & $V_{LSR}$ & Comment \\ 
name   & $\times 10^{-3}$ & (\kms)     &  (\kms)     & (K)   & (\kms)     &  (\kms) &  \\
\hline
G2.3+0.0 & 0.55(0.04) &  16.1(1.3)  &  6.4(0.6)  & 48.5(1.5) & 4.2(0.2) & 6.5(0.1) & R-C cloud \\
G4.7+0.0 & 0.76(0.06) &  8.5(0.07)  &  7.0(0.3)  & 64.7(2.3) & 4.7(0.2) & 7.0(0.1) & R-C cloud \\
G7.0+0.0 & 0.64(0.08) &  12.1(1.9)  &  8.0(0.8)  & 51.0(1.5) & 3.9(0.1) & 7.1(0.1) & R-C cloud\\
G13.9+0.0 & 0.35(0.04)$\:^{3}$&   6.8(1.0)  & 18.4(0.4)  & 10.4(0.8) & 5.3(0.5) & 19.3(0.2) & \\
G18.4+0.0 & 0.36(0.05)&  17.3(2.5)  & 24.1(1.1)  & 21.1(0.5)$\:^{4}$ & 6.0(0.2) & 22.8(0.1) & \\
\hline
\end{tabular}

\medskip
{$^1$ }{CRRL parameters for all positions except G13.9+0.0 are taken from Roshi \& Anantharamaiah (2001b).
The parameters for the position G13.9+0.0 are taken from Paper I. The CRRL data are obtained with an angular resolution
of $\sim$ 2\ddeg $\times$ 2\ddeg. \HI\ line parameters are obtained from Kalberla \etal\ (2005) after smoothing
the data to an angular resolution of $\sim$ 2\ddeg $\times$ 2\ddeg.} \\
{$^2$ }{$T_L/T_{C}$ is approximately the carbon line optical depth near 327 MHz. $T_{C}$ is the background continuum temperature
at the observed frequency.} \\
{$^3$ }{The given value is in $T_L/T_{sys}$, where $T_{sys}$ is the system temperature.} \\
{$^4$ }{The HISA parameters are obtained from the 0\ddeg.6 $\times$ 0\ddeg.6 \HI\ spectrum centred at $l = 18$\ddeg.5 and $b = 0$\ddeg.0.}
\end{minipage}
\end{table*}

We use the median line width ($\sim$ 17 \kms) to classify the CRRLs detected in the ORT survey 
into two groups -- (a) lines with width (FWHM) \lsim\ 17 \kms\ (``narrow lines'')
and (b) ``broad lines'' with width \gsim\ 17 \kms.  The median line width
is obtained from the 2\ddeg\ $\times$ 2\ddeg\ survey data.  
In Fig.~\ref{fig1}, we show an example CRRL spectrum, obtained toward $l = $ 13\ddeg.9, 
$b = $ 0\ddeg.  The spectrum shows
broad ($\Delta V $ = 41.5 \kms) and narrow ($\Delta V $ = 6.8 \kms) components. 
The broad lines detected toward G13.9+0.0 and other direction in the survey
may consist of several narrow components
as indicated by the higher angular resolution observations (see paper I). However, further
high sensitivity, high angular resolution observations are needed to confirm this
and hence we do not discuss the broad CRRLs any further.  
In this paper we focus on the origin of the narrow line emitting region. 
The parameters of the narrow line emission obtained from a Gaussian fit to
the spectrum towards  $l = $ 13\ddeg.9, $b = $ 0\ddeg~ are given in Table~\ref{tab1}.
The \HI\ spectrum toward the same direction obtained from
the Leiden/Argentine/Bonn (LAB) survey (\nocite{ketal05}Kalberla \etal\ 2005, angular resolution 
0\ddeg.6 $\times$ 0\ddeg.6) is shown in the middle panel of Fig.~\ref{fig1}. The spectrum 
obtained after smoothing the LAB data to $\sim$2\ddeg\ $\times$  2\ddeg\ resolution is 
shown in the lowermost panels of Fig.~\ref{fig1}.  
A HISA feature is seen at the same LSR (Local Standard of Rest)
velocity as the narrow carbon line. 
The HISA is prominent in the higher resolution 
(0\ddeg.6 $\times$ 0\ddeg.6) spectrum (see Fig.~\ref{fig1}). 
The parameters of the HISA are included in Table~\ref{tab1}. 
The coincidence of the LSR velocities of the two lines implies 
that the two line forming regions coexist,
if we make the standard assumption that the LSR
velocities are due to galactic differential rotation; 
spectral traces having the same LSR velocities originate from regions
at the same line-of-sight (LOS) distances.
The widths of the two spectral lines,
CRRL and \HI\ line, differ with the CRRL being broader compared to \HI\ line.
The origin of this difference is discussed in Section~\ref{rccloud}.

Majority (63\%) of the narrow CRRLs observed in the 2\ddeg\ $\times$ 2\ddeg\ ORT survey
have a corresponding HISA feature at the same LSR velocity.
Absence of a corresponding HISA in some of the directions where 
narrow CRRLs are observed may be due to the following reason. 
Detection of HISA depends on favourable
observing conditions. In order to detect a cool \HI\ cloud in 
self-absorption, background \HI\ emission with brightness
temperature greater than the spin temperature of the cool cloud
is required. Variation in the background emission temperature
over the observing region can make the self-absorption difficult
to detect, especially when observed with a coarse angular resolution.
The need for higher angular resolution \HI\ observations  to
detect HISA has been demonstrated, for example, by  
\nocite{bl84}Bania \& Lockman (1984).  CRRLs 
do not need such favourable conditions for their detection. 
In Fig.~\ref{fig2} we show examples where narrow carbon lines are detected but no 
HISA is seen in the 2\ddeg\ $\times$ 2\ddeg\ averaged \HI\ spectrum (lowermost panels). 
While the higher angular resolution (0.6\ddeg\ $\times$ 0.6\ddeg) 
\HI\ spectrum (middle panel) shows an HISA feature towards G18.4+0.0, 
no such feature is seen toward G16.1+0.0 
even in the higher angular resolution spectrum. 

The \HI\ spectra obtained close to the Galactic plane show a
wealth of structures and many of these structures are 
self-absorption features as shown, for example, by \nocite{bl84}Bania \& 
Lockman (1984) in their  high angular resolution observations. The
ORT survey have not detected carbon lines corresponding to
all these features. The typical upper limit on the CRRL optical
depths from these features is $\sim$ 2.0 $\times 10^{-4}$
(Roshi \etal\ 2002). 
The possible reasons for these are : (1)
the CRRL survey is biased toward detecting carbon line emitting
regions with large angular extent such that the beam dilution 
factor is insignificant; (2) the optical depth of CRRLs
in all the regions are not high enough to detect them.
Variation in carbon optical depth is seen, for 
example, towards the R-C cloud (see Section~\ref{rccloud}).  
(3) the Galactic non-thermal background radiation field is not
strong enough to make the CRRL detection possible. The low
Galactic radiation field may be the reason for non-detection of
CRRLs toward HISA at $l$ \gsim 20\ddeg.   

Difficulties in quantifying the observed properties of HISA 
features have been elaborated by several
authors (eg. \nocite{lb80}Levinson \& Brown 1980). The observed line parameters
listed in Table~\ref{tab1} are obtained by fitting
a Gaussian to the absorption feature after removing  
2nd or 3rd order polynomial ``baseline'' to the
\HI\ emission near this feature. \nocite{lb80}Levinson \& Brown (1980)
have also shown, through simulation, that the observed central velocity of
the \HI\ absorption can be `shifted' compared to the actual central velocity
and this shift depends on the slope of the background \HI\ emission. A
rough estimate of this shift in the cases listed in Table~\ref{tab1}
shows that it is insignificant compared to the errors in the line parameters. 
 
A well known HISA feature towards the Galactic centre direction is the 
Riegel-Crutcher (R-C) cloud. In the next section, we concentrate 
on CRRL detection toward the R-C cloud. In particular, we demonstrate
the usefulness of combining CRRL and \HI\ observations to infer the
physical properties and processes in the cloud. The physical processes
that are discussed here are now part of well known numerical codes, which implement
models for Photodissociation region (PDR; see for example \nocite{ht97}Hollenbach \& Tielens 1997,
\nocite{htt91}Hollenbach \etal\ 1991).
Here we present a semi-analytic estimation of the physical properties
and energetics in the R-C cloud. A detailed PDR modelling will be presented elsewhere.

\section{The Riegel-Crutcher Cloud}
\label{rccloud}

A prominent cool neutral cloud (the R-C cloud; \nocite{rc72}Riegel \& Crutcher 1972, 
\nocite{h55}Heeschen 1955) has been observed in \HI\ self-absorption  
in early surveys of the Galactic centre region. 
The self-absorption cloud has an extent of
at least 40\ddeg\ along the galactic longitude and $\sim$ 10\ddeg\ along galactic
latitude (\nocite{rc72}Riegel \& Crutcher 1972, \nocite{rj69}Riegel \& Jennings 1969). 
Line emissions from molecules such as \tCO\ and OH have been detected
in many directions toward the R-C cloud (\nocite{c73}Crutcher 1973). 
Distance to the R-C cloud was determined to be 125$\pm$25 pc from
\NaI\ observations against background stars (\nocite{cl84}Crutcher \& Lien 1984). 
Optical observations have
also constrained the LOS thickness of the cloud
to be between 1 and 5 pc (\nocite{cr74}Crutcher \& Riegel 1974). 
Recent high resolution ($\sim$ 100\arcsec) 
\HI\ line observations have revealed filamentary structures in the cloud with typical
transverse width of 0.1 pc (\nocite{metal06}McClure-Griffiths \etal\ 2006). 
\nocite{metal06}McClure-Griffiths \etal\ (2006) also discussed the possibility that 
the LOS extent of the R-C cloud may be much smaller than 5 pc inferred by
\nocite{cr74}Crutcher \& Riegel (1974) from their optical observations. They suggested that 
the LOS thickness may be $\sim$ 0.1 pc, similar to the transverse width of the filaments.
The \HI\ absorption measurements have been used to infer a mean spin temperature of $\sim$ 40 K
and \HI\ column density \NHI\ of $\sim$ 
10$^{20}$ \cmtwo\ for the R-C cloud (\nocite{mbd95}Montgomery, Bates, Davies 1995, 
\nocite{metal06}McClure-Griffiths \etal\ 2006).  

\begin{figure*}
\includegraphics[width=3.5in, angle = -90]{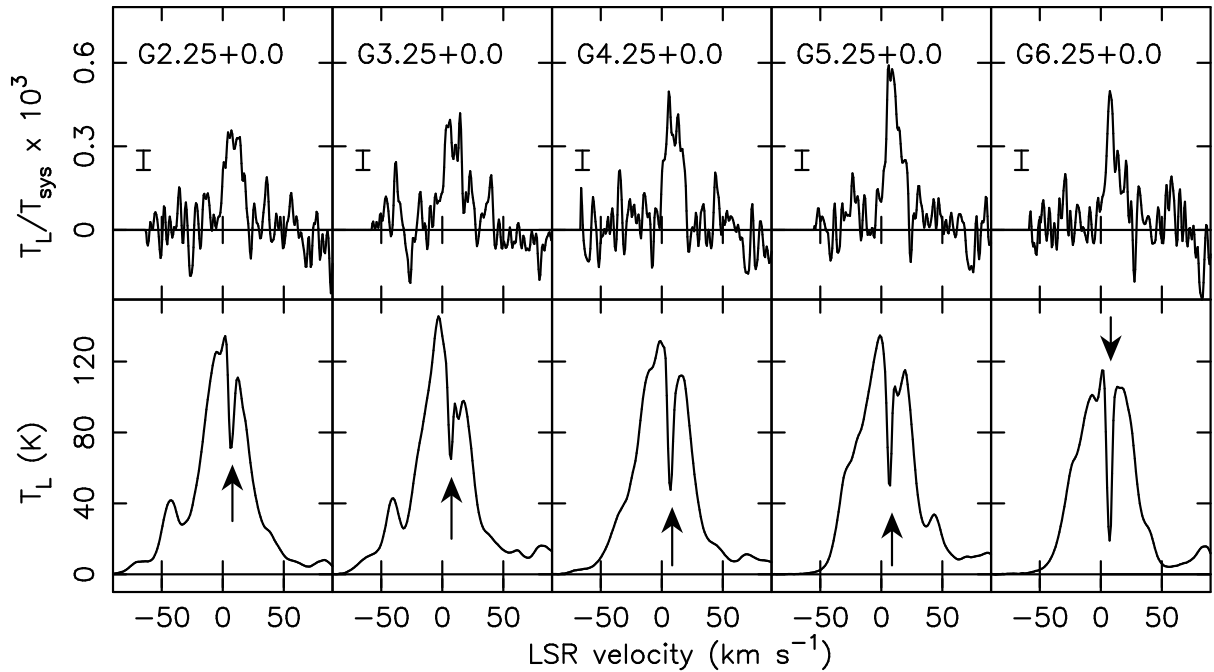}
\caption{CRRL (top panels) and \HI\ (bottom panels) spectra toward the Riegel-Crutcher cloud.
CRRL spectra at 327 MHz and \HI\ spectra (Kalberla \etal\ 2005) are obtained with 
angular resolution of 2\ddeg\ (along $b$) 
$\times$ 0\ddeg.5 (along $l$) and 0\ddeg.6 $\times$ 0\ddeg.6 respectively 
The 1$\sigma$ values of the noise in CRRL spectra are also shown in the top panels. 
The spectra are centred near the Galactic coordinates marked in the top panel.
The arrows are placed at the LSR velocity of the
carbon lines detected at the different positions. 
}
\label{fig3}
\end{figure*}

\HI\ spectra toward the R-C cloud in the longitude range $\sim$ 2\ddeg\ to 7\ddeg\
are shown along with the CRRL spectra in Fig.~\ref{fig3}. The angular resolution
of \HI\ spectra is 0\ddeg.6 $\times$ 0\ddeg.6 and that of CRRL spectra is 
2\ddeg\ (along $b$) $\times$ 0\ddeg.5 (along $l$). The prominent self-absorption feature
seen in the \HI\ spectra is due to the R-C Cloud. The \HI\ line and CRRL parameters
obtained from the spectra with 2\ddeg\ (along $b$) $\times$ 2\ddeg\ (along $l$) resolution 
in the same longitude range are given in Table~\ref{tab1}. 
It is evident from Fig.~\ref{fig3} and the line parameters given in Table~\ref{tab1} that the 
LSR velocities of CRRLs and HISA coincide.  Based on this coincidence 
we conclude that the carbon line forming regions are associated with the R-C cloud.

The R-C cloud extends from $\sim -15^{\circ} < l < \sim25^{\circ}$ 
(\nocite{rc72}Riegel \& Crutcher 1972). However, CRRL emission from the R-C cloud 
could be identified only in the longitude range $\sim$ 2\ddeg\ to 7\ddeg. Close to the
Galactic centre (ie $|l|$ \lsim\ 2\ddeg) it is difficult to identify the CRRL emission
associated with the R-C Cloud due to the degeneracy in LSR velocity of recombination line emission 
from several regions along the LOS. In other longitude range 
spanned by the R-C cloud the non-detection of
CRRL emission may be either due to sensitivity limitation or due to variation in fraction of
ionised carbon in the R-C cloud. 

\subsection{CRRL and HISA line widths}
\label{lwidth}

The average width of the carbon line from the R-C cloud  is $\sim$ 12 \kms\ (FWHM) which is 
about 3 times the width of the \HI\ absorption line (FWHM $\sim$ 4.0 \kms).
CRRLs detected in several other directions also have a larger width compared to the \HI\ line 
(see Table~\ref{tab1}). If the two line forming regions coexist then
the spectral lines from such regions are expected to have the same widths
(this is true if the line widths are dominated by non-thermal motions, which
is the case for the R-C cloud). 
A possible explanation for the difference in the line widths of CRRL and
\HI\ is the following.  As mentioned above, detection of cool \HI\ clouds 
in self-absorption needs favourable
observing conditions and angular resolution (\nocite{bl84}Bania \& Lockman 1984).
On the other hand, detection of CRRLs does not need such favourable conditions as long as
the line forming regions fill a substantial portion of the observing beam and the Galactic
background radiation field is strong.  If there are several cool \HI\ clouds
with different velocities within the 2\ddeg\ $\times$ 2\ddeg\ field
of the CRRL observations, the coarser beam will detect a
broad carbon line while the \HI\ spectrum will
be dominated by absorption due to the coldest gas. 
In addition, any velocity gradient within the observing beam
can also contribute to the line width. Such a velocity gradient
has been detected for \HI\ absorption in the R-C cloud
(\nocite{mbd95}Montgomery \etal\ 1995). The R-C cloud also exhibits
multiple \HI\ absorption features (eg. \nocite{mbd95}Montgomery \etal\ 1995).
Comparing CRRL and \HI\ spectra obtained with similar (high) angular 
resolution can help in evaluating these possibilities.

\subsection{Modelling the line forming region in the R-C cloud}
\label{model}

\begin{table}
\small
\caption{CRRL parameters observed toward the R-C Cloud}
\label{tab2}
\begin{tabular}{lrrccr}
\hline
Freq & CRRL$\:^{1}$         & $\int \tau_L d\nu\:^{2}$ & $\Delta V\:^{2}$  & $V_{LSR}\:^{2}$ & Ref  \\ 
(MHz)&              & (s$^{-1}$)         &  (\kms)     & (\kms)    &      \\
\hline
327  & C272$\alpha$ & 9.2(1.4)           & 12.2(1.4)   &  7.1(0.6) & 1  \\
76   & C441$\alpha$ & $-$3.9(0.7)        & 18.4(2.6)   &  5.5(1.3) & 2  \\
76   & C555$\beta$  & $-$3.3(0.6)        & 20.5(3.1)   &  2.2(1.6) & 2  \\
34.5 & C575$\alpha$ & $-$1.9(0.3)        & 21.2(2.0)   & 10.2(1.4) & 3 \\
\hline
\end{tabular}

\medskip
{$^1$ }{Multiple transitions were observed at all three frequency bands. We list
here the central transitions observed at each band.} \\
{$^2$ }{The 1$\sigma$ errors on the line parameters are given in bracketed values}  \\\\
{\em References: } (1)Roshi \& Anantharamaiah (2001b); (2) Erickson \etal\ (1995) ;
(3) Kantharia \& Anantharamaiah (2001)
\end{table}

The CRRL data toward the R-C cloud at 327 MHz combined with existing data at
76 MHz (angular resolution $\sim$ 4\ddeg; \nocite{ema95}Erickson \etal\ 1995) 
and 34.5 MHz (angular resolution $\sim$  21\arcmin $\times$ 25\ddeg; 
\nocite{ka01}Kantharia \& Anantharamaiah 2001) 
are used to model the physical properties of the line forming region.
At 76 MHz both $\alpha$ and $\beta$ transitions were detected toward the R-C cloud.
To study the average properties of the \CII\ region in the R-C
cloud, all available CRRL data in the range $l \sim 2$\ddeg\
to $\sim 7$\ddeg\ and $b \pm 2$\ddeg\ at 
327 and 76 MHz were averaged.
These averaged values are listed in Table~\ref{tab2}.
Since the 34.5 MHz observations used a fan beam, no averaging was done and
we have included the parameter fit to the spectrum towards $l=5^{\circ}$ and $b=0^{\circ}$. 
The line width and central velocities 
for all these transitions roughly match; any differences,
particularly the difference between the 327 and 34.5 MHz line parameters, 
are attributed to the differences in the observing beam.     
To determine the physical properties, we followed the method described by 
\nocite{ka01}Kantharia \& Anantharamaiah (2001) where a uniform slab of
line emitting region with electron temperature, $T_e$, electron density, $n_e$, and LOS 
extent, $S$, is considered.

The integrated line optical
depth is  related to these parameters through the equation (\nocite{s75}Shaver 1975) 
\be
\int \tau_L d\nu \approx 1.07 \times 10^7\; b_n\; \beta_{n,\Delta n}\; K(\Delta n)\; \Delta n\; T_e^{-2.5} n_e^2\; S \mbox{~~s}^{-1},
\ee
where $b_n$ and $\beta_{n,\Delta n}$ are the non-LTE departure coefficients for principal quantum number $n$ and
transition $\Delta n$. K($\Delta n$) = 0.1908 and 0.02633 for $\Delta n$ = 1 and $\Delta n$ = 2 
transitions respectively.  $\beta_{n,\Delta n}$ is defined as
\be 
 \beta_{n,\Delta n} = 1 - 3.2 \times 10^{-6} \frac{n^3}{\Delta n} T_e \frac{b_n - b_{n+\Delta n}}{b_n}.
\ee
In the above equations the units of $T_e$ is K, $n_e$ is \cmthree and S is pc.
The departure coefficients, which are computed using the programs of 
\nocite{pae94}Payne \etal\ (1994), a modified version of the original
program of \nocite{ww82}Walmsley \& Watson (1982), 
depend on the background continuum radiation field.  We used 5000 K at 100 MHz 
as the background temperature. This background temperature is obtained from 
the continuum map at 34.5 MHz (\nocite{d89}Dwarakanath 1989; see 
also \nocite{ka01}Kantharia
\& Anantharamaiah 2001) and scaled to
100 MHz using a spectral index of $-$2.6. 
The derived parameters of the line emitting region change
by a few percent for a factor of 2 change in the 
background temperature. 
An abundance for carbon A$_c$ of 1.4 $\times$ $10^{-4}$ 
obtained from the solar abundance of 2.9 $\times$ $10^{-4}$ 
(\nocite{l03}Lodders 2003) and assuming a depletion factor of 
0.48 (\nocite{j09}Jenkins 2009, \nocite{wetal03}Wolfire \etal\ 2003)  
is used for the modelling.
Since the R-C cloud has a large angular extent, no beam dilution factor 
is used for the 327 and 76 MHz observations
to convert the observed line antenna temperature to brightness temperature. 
We could not find a single model which fitted all the three observed points.
The 34.5 MHz observations were made with a beam $\sim$ 25\ddeg\ in size 
along Galactic latitude and hence the observed line optical depth may have
to be corrected by an unknown beam dilution factor.
Therefore we constrained the model parameters using the 76 and 327 MHz data. 
Modelling of the data at
76 and 327 MHz resulted in the line forming regions having the following
physical properties:  $T_e$ $\sim$ 40 $\rightarrow$ 60 K, $n_e$ $\sim$
0.8 $\rightarrow$ 0.05 \cmthree and $S$ $\sim$ 0.03 $\rightarrow$ 3.5 pc. 
For $T_e$ larger than 60 K we find that models with lower $n_e$ (\lsim 0.05 \cmthree) 
also fit the CRRL data, however, the path lengths are longer than the LOS extent
of 5 pc which is the thickness of the RC cloud as obtained by \nocite{cr74}Crutcher \& Riegel (1974).
We, hence, rule out these higher temperature models.  

\begin{table*}
\small
\centering
\begin{minipage}{140mm}
\caption{Physical properties of the Riegel-Crutcher cloud$^a$}
\label{tab3}
\begin{tabular}{cccccccccccc}
\hline
Model & $T_e$ & $n_e$        & S   &  n$_H$     & Av & \nHI & n$_{H_2}$  & $f$  & $P_{HI}\:^1$        & \NHI     & \NHtwo \\ 
No.   & (K)   & (\cmthree)   & (pc)&  (\cmthree)&  & (\cmthree) & (\cmthree) &      &  & (\cmtwo) & (\cmtwo) \\
      &       &              &     &           &   &     &      &      & ($\times 10^3$)& ($\times 10^{20}$)& ($\times 10^{20}$) \\ 
\hline
1     & 60    & 0.1      & 1.1     & 700     & 1.4 & 90 & 300  & 0.9  &  5           &  3       & 11  \\ 
2     & 50    & 0.3      & 0.15    & 2100    & 0.6 & 500& 800  & 0.8  &  27          &  2.5     & 3.7 \\
3     & 40    & 0.6      & 0.04    & 4300    & 0.3 & 1430& 1430 & 0.7  &  57          &  1.8     & 1.8 \\
\hline
\end{tabular}

\medskip
{$^a$ }{$T_e$, $n_e$, $S$ are
the model parameters, \nH\ is the hydrogen 
nuclear density, Av is the visual extinction, \nHI\ is the hydrogen atomic density, \nHtwo\
is the hydrogen molecular density, $f = 2$\nHtwo/\nH\ is the molecular fraction, \NHI\ and
\NHtwo\ are the atomic and molecular column densities respectively. }

{$^1$ }{Partial pressure of \HI\ is tabulated in units of K \cmthree.}
\end{minipage}
\end{table*}

\begin{figure*}
\includegraphics[width=3.5in, angle = -90]{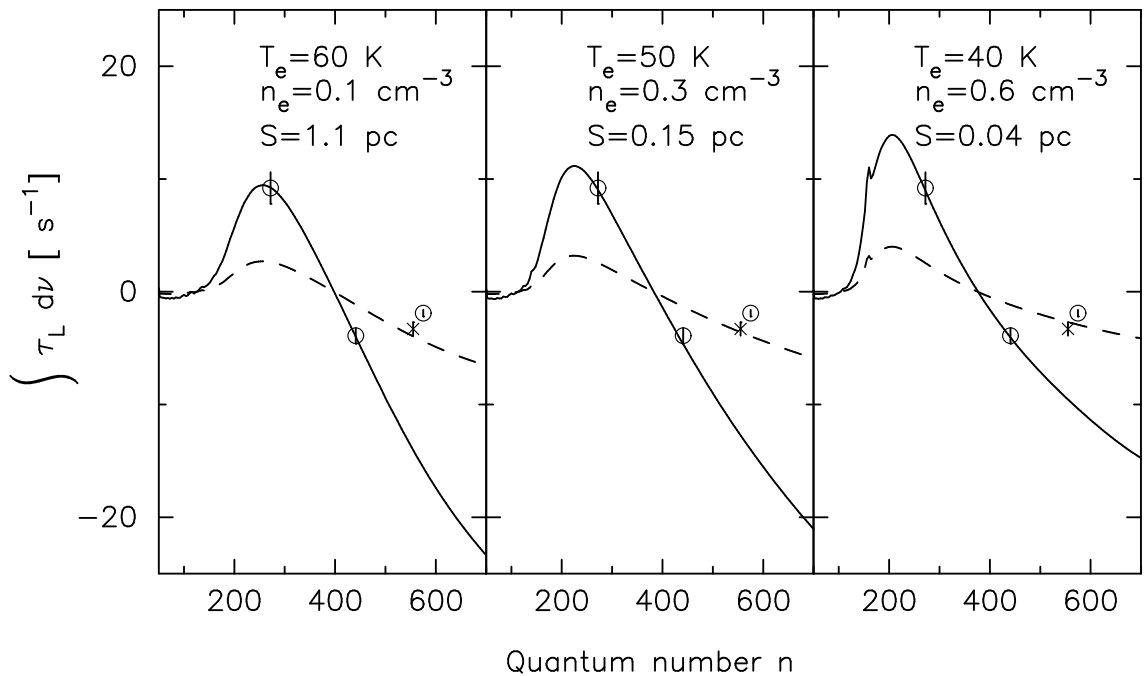}
\caption{Integrated CRRL optical depth from models for the Riegel-Crutcher cloud plotted against 
principal quantum number.  
The optical depths of Cn$\alpha$ and Cn$\beta$ 
for the best fit model parameters 
are shown by solid and dashed lines respectively. 
The value of model parameters (electron temperature, $T_e$, electron
density, $n_e$, and line-of-sight path length, $S$) are indicated on the plots. 
The observed integrated optical depths of Cn$\alpha$ lines are indicated by circles  and that of
Cn$\beta$ line is shown by cross. The error bars represent $\pm$ 1$\sigma$ values.
}
\label{fig4}
\end{figure*}

The above modelling for the carbon lines used the departure coefficients estimated after including
the dielectronic-like recombination process 
(\nocite{wwc80}Watson, Western \& Christensen 1980, \nocite{ww82}Walmsley \& Watson 1982, 
see also \nocite{pae94}Payne \etal\ 1994) which involves the excitation of
the fine structure transition in the core electrons in carbon giving rise to a spectral line at 
$158 \mu$m. This process depends on the
departure coefficients (RTE; \nocite{ww82}Walmsley \& Watson 1982) of the 2P$_{3/2}$ and 
2P$_{1/2}$ states in carbon.  RTE is obtained using
statistical equilibrium for the 2P$_{3/2}$ and 2P$_{1/2}$ states.
The equation of statistical equilibrium includes the density of colliding particles 
(electrons, H and H$_2$) which are estimated as described in Subsection~\ref{atommol},
and iterated to get a consistent model.
Table~\ref{tab3} gives three representative models consistent with
the 327 and 76 MHz (both $\alpha$ and $\beta$ transitions) CRRL observations. 
The expected integrated line optical depth as a function of quantum
number for the three sets of model parameters listed in Table~\ref{tab3}  
are shown in Fig.~\ref{fig4}. For each temperature listed in Table~\ref{tab3}, 
the models are consistent with the data for a factor of 2 and 4 change
in the listed densities and path lengths respectively.  

\subsection{Atomic and molecular hydrogen densities}
\label{atommol}

We combine the CRRL and \HI\ data to get more insight into the physical
state of the R-C cloud.
The inferred \NHI\ for R-C cloud from \HI\ observations
is between a few times 10$^{19}$ to 4 $\times$ 10$^{20}$ \cmtwo  
(\nocite{mbd95}Montgomery \etal\ 1995, \nocite{metal06}McClure-Griffiths \etal\ 2006)
which is typical of CNM clouds in our Galaxy. 
This high \NHI\ implies that 
photoionisation due to EUV/soft X-ray does not dominate in the interior of the R-C cloud 
(see, for example, \nocite{gl74}Glassgold and Langer 1974) and most of the ionisation
is due to FUV photons which ionise carbon atoms. 
So, unlike in CNM clouds with canonical ISM pressure ($\sim$ 3000 K \cmthree), 
where electrons are due to ionisation of hydrogen atoms (Wolfire \etal\ 2003), 
most of the 
electrons in the R-C cloud are due to carbon ionisation.  This fact can thus
be used to estimate the number density of hydrogen, 
n$_H$, in the cloud; n$_H$ = n$_e$ / A$_c$ 
(see Table~\ref{tab3}) where A$_c$ is the number abundance of gas phase carbon atoms 
taken to be $1.4\times10^{-4}$ (see Subsection~\ref{model}).  
The fraction of hydrogen which is tied up in the atomic
form, ie \HI , is inferred as follows. 
The \HI\ optical depth from the CRRL models is obtained by assuming that
(a) some fraction of n$_H$ is in atomic form (the remaining fraction 
is H$_2$ molecules) and (b)  the
spin temperature is equal to the electron temperature, which is generally 
the case for cold \HI\ regions (eg. \nocite{kh88}Kulkarni \& Heiles 1988).
In addition, we assume that the carbon and \HI\ line forming regions co-exist 
along the LOS. This assumption means that, for models with $T_e \sim$ 60 K and 50 K 
the LOS thickness of the \HI\ region is respectively $\sim$ 1.1 pc, close to the value suggested
by \nocite{cr74}Crutcher \& Riegel (1974; ie 1 -- 5 pc), and $\sim$ 0.15 pc, close 
to the suggested value of 0.1 pc by McClure-Griffiths \etal\ (2006). 
For models with $T_e \sim$ 40 K, the LOS thickness of the \HI\
region will be 0.05 pc. 
For these models, it is possible that the CRRL emission originates from an interface
region between the \HI\ and \Htwo\ in the R-C cloud. PDR modelling of the
R-C cloud is needed to investigate this possibility. 
Following literature, we define the molecular fraction in terms of the \nHtwo\
content; $f = \frac{2 n_{H_2}}{n_{HI} + 2 n_{H_2}} = \frac{2 n_{H_2}}{n_{H}}$,
where $n_{H_2}$ and $n_{HI}$ are the H$_2$ and \HI\ densities respectively.  
The \HI\ optical depth due to CRRL models are then equated to the observed value 
(mean peak optical depth in the R-C cloud $\sim$ 0.7; 
McClure-Griffiths \etal\ 2006) to determine $f$. 
The estimated $f$, \nHI, \nHtwo\ and \HI\ partial pressure, 
$P_{HI}$ = \nHI $T_e$ 
for the three representative models are listed in Table~\ref{tab3}.

The higher temperature models (ie $T_e$ $\sim$ 60 K) have relatively low electron
and \HI\ densities. Their LOS extent is about a pc or more. The partial pressure of hydrogen
is close to the mean interstellar pressure in the Solar neighbourhood. The molecular
density compared to the \HI\ density is higher in these models. 
The LOS extent of lower temperature models (ie $T_e$ $\sim$ 40 K) 
is $\sim$ 0.05 pc. They have larger \HI\ and electron densities.
The molecular density is about the same as that of \HI\ density. The \HI\ 
partial pressure in these models is about an order of magnitude higher than the mean interstellar
pressure. Note that for all the models the total gas pressure is at least an
order of magnitude higher than the mean interstellar pressure and hence the cloud
has to be supported by either gravity or magnetic pressure. In the next subsection,
we compare the model predictions with other existing UV and optical observations
with the intention of narrowing down the range of parameter values. 

\subsection{Comparison with UV and Optical observations}

Several stars beyond the R-C cloud have been observed in the optical and UV, thus
sampling the gas in the intervening cloud.  From these observations we take 
data toward two stars, HD165246 (l=6\ddeg.4, b=$-$1\ddeg.56, distance=1.85 kpc;
\nocite{j09}Jenkins 2009) \&  HD164402 
(l=7\ddeg,b=0\ddeg, distance=1.74 kpc; \nocite{setal77}Savage \etal\ 1977), 
which overlap with the 
directions in which CRRLs from the R-C cloud are observed. The measured visual extinction Av, towards 
these two stars are 1.1 and 0.95 and the measured \NHtwo\ are 
1.4 $\times 10^{20}$ \cmtwo\ and 3 $\times 10^{19}$ \cmtwo 
(\nocite{j09}Jenkins 2009, \nocite{setal77}Savage \etal\ 1977) respectively.
About 10\% of the extinction and almost all the H$_2$ are due to the R-C cloud (Montgomery \etal\ 1995). 
Av from our model parameters using the equation Av = NH/($1.7 \times 10^{21}$) 
(\nocite{bsd78}Bohlin, Savage \& Drake 1978)
are listed in Table~\ref{tab3}.  The observed extinction due to the R-C cloud ie 0.11 to 0.095 is 
closest to our lower temperature ($T_e \sim 40$ K) model which predicts an extinction
of 0.3 and \NHtwo\ of $1.8\times10^{20}$ \cmtwo.   
Spectroscopic observations of the spectral lines of \NaI\ (\nocite{cr74}Crutcher \& Riegel 1974), 
\MgI\ and \MgII\ (\nocite{bmk95}Bates, Montgomery, Kemp 1995) toward stars 
behind the R-C cloud can be used to determine the electron density in the cloud. The
derived electron densities are typically $<$ 0.1 \cmthree.   Although we did find
models with such low electron densities which fitted the observed data points, we do
not favour these due to the long path-lengths, and hence correspondingly high Av, required 
to explain the observed line
strengths.  However, relaxing some of the assumptions made in deriving $n_e$ from optical
line observations can increase the estimated electron density. For example,
\nocite{bmk95}Bates \etal\ (1995) obtained $n_e \sim$ 0.3 \cmthree 
by considering  that the \MgI\ line is mainly
from the cloud core and the \MgII\ is distributed along the line of sight.  
This value is within the range of
models that we derive for the carbon line forming region in the R-C cloud 
(see Table~\ref{tab3}).  However it is not sufficient to favour the lowest
temperature model over the others. Thus, it appears that it is difficult to
narrow down the range of physical parameters listed in Table~\ref{tab3}
for the R-C cloud with existing data.  

\section{The physical state of the R-C Cloud}
\label{hcm}
In this section, we use the model parameters to investigate the 
cooling and heating processes in the R-C cloud. We also estimate
the neutral carbon fraction and molecular formation and dissociation
rates in the cloud. 

\subsection{Cooling in the R-C cloud}

\begin{table*}
\small
\centering
\begin{minipage}{140mm}
\caption{Thermal properties and $H_2$ formation/dissociation in the R-C cloud$^a$}
\label{tab4}
\begin{tabular}{cccccccc}
\hline
Model &  $\Lambda_{CII} ^{\:1}$ & $\Lambda_{other}^{\:1}$         & $\int I_{CII} d\nu^{\:2}$ &  G0 & $\Gamma_{phdiss}^{\:1,3}$ & $R_{form}$  & $R_{diss}^{\:3}$ \\ 
No.   &  &  &    &      &  & (\cmthree s$^{-1}$) & (\cmthree s$^{-1}$)  \\
      & ($\times 10^{-23}$) & ($\times 10^{-23}$) & ($\times 10^{-5}$) &    &  ($\times 10^{-25}$)        & ($\times 10^{-11}$) & ($\times 10^{-11}$) \\ 
\hline
1     &  12 & 3        & 3.3    & 4 & 0.1   & 0.2 & 0.003 \\ 
2     &  74 & 8        & 2.7    & 7 & 5.6   & 2.4 & 0.14 \\
3     &  153 & 14    & 1.5    & 6 & 28    & 11.6 & 0.7 \\
\hline
\end{tabular}

\medskip
{$^a$ }{ $\Lambda_{CII}$ is the cooling rate due to \CII\ 158 \mum\ radiation, $\Lambda_{other}$
is the total cooling rate due to atomic and molecular line emission, $I_{CII}$ is the
intensity of \CII\ 158 \mum\ line emission, G0 is the flux density of FUV radiation field in Habing
unit, $\Gamma_{phdiss}$ is the heating rate due to the dissociation of \Htwo\ molecules,
$R_{form}$  and $R_{diss}$ are the \Htwo\ formation and dissociation rate respectively.}

{$^1$ }{The units of $\Lambda_{CII}$, $\Lambda_{other}$ and $\Gamma_{phdiss}$ are ergs s$^{-1}$ \cmthree} \\
{$^2$ }{The intensity of \CII\ line is tabulate in units of ergs s$^{-1}$ \cmtwo sr$^{-1}$} \\ 
{$^3$ }{$\Gamma_{phdiss}$ and $R_{diss}$ are estimated at Av/2. The Av obtained for the models is given in Table 3.}
\end{minipage}
\end{table*}

The derived properties of the R-C cloud are used to determine the
cooling rate in the gas. The major cooling processes in these clouds 
are due to transitions in \CII, \CI\ and \OI\ and molecular
transitions in \Htwo\ and CO.  The \CII\ 158 \mum\ is believed to
be the major coolant in diffuse clouds with temperatures $\sim$ 100 K
(\nocite{dm72}Dalgarno \& McCray 1972). The cooling rate due to
\CII\ 158 \mum\ line emission is calculated following 
\nocite{w84}Watson (1984) and using the values for 
collision rate coefficients and Einstein A-coefficient given 
by \nocite{setal05}Sch\"{o}ier \etal\ (2005). The combined cooling
rate due to \CI\, \OI\ and molecular transitions in \Htwo\ and CO
is calculated from Fig 1. of \nocite{g84}Gilden (1984). These
cooling rates along with the estimated intensity of the \CII\ 158 \mum\
line are given in Table~\ref{tab4} for the three representative models. 
The combined cooling rate due to atoms and molecules is at
least a factor of 4 smaller than that due to the \CII\ 158 \mum emission. 

\subsection{Heating in the R-C cloud}

The heating processes which are important in diffuse clouds and
considered here are photoelectric emission, H$_2$ dissociation,
carbon ionisation and cosmic rays. 
The heating efficiency of photoelectric emission depends on the
grain charge which in turn is a function of the electron density and G0,
the interstellar FUV (6 to 13.6 eV) radiation field in Habing unit 
(1.6 $\times 10^{-3}$ ergs sec$^{-1}$ \cmtwo; \nocite{h68}Habing 1968). 
The heating rate per unit volume due to photoelectric emission is
given by (\nocite{wetal03}Wolfire \etal\ 2003) 
\be
\Gamma_{pe} = 1.3 \times 10^{-24}\; n_H\; \epsilon\; G0\:  \mbox{ ergs s}^{-1} \mbox{ cm}^{-3},
\ee
where the photoelectric emission efficiency, $\epsilon$ is 
\bea
\epsilon = & \frac{4.9\times 10^{-2}}{1+4\times 10^{-3}(G0\; T_e^{0.5}/(n_e \phi_{PAH})^{0.73})} + \nonumber \\
           & \frac{3.7\times 10^{-2}(T_e/10^{4})^{0.7}}{1+2\times 10^{-4}(G0\; T_e^{0.5}/n_e \phi_{PAH})}.
\eea
In the above equation, $\phi_{PAH}$ is a parameter introduced by Wolfire \etal\ (2003) to modify
the electron-dust collision rates; following them
we take its value to be 0.5. The values for $n_e$, \nH\ and $T_e$ are taken from Table~\ref{tab3}
for estimating the heating rate. 

The second process we examine is fluorescent photodissociation of  
H$_2$.  This process results in energetic H atoms which in turn lead to the heating of the cloud. 
The heating rate is essentially  
the product of the photodissociation rate per unit volume (see
Eq.~\ref{eqphdis} in Subsection~\ref{formdiss}) and the mean kinetic energy ($\sim$ 0.25 eV)
of the dissociated atoms (\nocite{sd73}Stephens \& Dalgarno 1973; \nocite{t05}Tielens 2005). 
We examine the relative importance of the four processes in heating the HISA cloud
for G0 ranging from 1 to 10. 
We find that heating due to carbon ionisation (\nocite{t05}Tielens 2005) for a carbon neutral fraction
\lsim 0.08 (see Subsection~\ref{neuc}) as well as cosmic ray heating are 
insignificant compared to the other two heating processes. 

The photodissociation
heating depends on the dissociation rate $R_{phdiss}$, which is a function of
FUV radiation intensity inside the cloud. As described in 
Subsection~\ref{formdiss}, opacity of FUV lines in the cloud plays an
important role in determining the dissociation rate inside the cloud.
It can be shown that the opacity effect reduces the dissociation
rate considerably at \Htwo\ column densities $> 10^{14}$ \cmtwo. This
effect is termed `self-shielding' 
(see for example \nocite{db96}Draine \& Bertoldi 1996). Observations show that 
the \Htwo\ column density of R-C cloud is $> 10^{19}$ \cmtwo  
(\nocite{j09}Jenkins 2009, \nocite{setal77}Savage \etal\ 1977). 
In clouds with such column densities detailed modelling shows 
that a gradient in the density ratio of hydrogen in atomic and molecular form 
exists and H$_2$ self-shielding becomes important (eg. \nocite{vb86}van Dishoeck \& Black 1986). 
Such detailed modelling, which are implemented in PDR codes
(for example \nocite{ht97}Hollenbach \& Tielens 1997),
are beyond the scope of the present work and will be presented
elsewhere. In the subsequent part of the paper, we provide
estimates of various quantities at a depth where the visual extinction (Av) is about half the
total Av due to the cloud.  We refer to this depth in the cloud as Av/2 and note that
self-shielding effects need to be included while estimating the physical processes in
the R-C cloud.
Estimates of the heating at a depth of Av/2 for G0 ranging between 1 and 10 shows that  
photoelectric heating dominates in the cloud interior.

As mentioned above, photoelectric and photodissociation heating depends on the background
FUV flux G0. Constraints on G0 may be obtained by assuming that the R-C cloud is in
thermal equilibrium ie by equating \CII\ 158 \mum cooling rate per unit volume 
to the heating rate per unit volume.  We estimate that 
G0 is between 4 and 7 (see Table~\ref{tab4}). For comparison with photoelectric
heating rate, which is
approximately equal to the $\Lambda_{CII}$ listed in Table~\ref{tab4}, the heating rates due to
photodissociation processes at Av/2 are included in Table~\ref{tab4} for the estimated G0. 

\subsection{Neutral carbon in the R-C cloud}
\label{neuc}
In this section, we estimate the neutral fraction of carbon 
in the R-C cloud using our model parameters. 
To a large extent, the background FUV flux and the fraction of neutral carbon 
determines the ionisation of carbon in the cloud
(see \nocite{gl75}Glassgold \& Langer 1975 for other factors affecting carbon ionisation). 
We estimate the neutral fraction by assuming that carbon ionisation is dominated
by FUV radiation and that all electrons are due to carbon ions. 
The ionisation equilibrium of carbon implies 
\be
n_e n_{C^+} \alpha_R = f_C n_{C^+} \Gamma_{ion},
\label{eqcion}
\ee
where $n_{C^+}$ is the number density of carbon ions,
$\alpha_R = 6.38 \times 10^{-11}$ cm$^3$ s$^{-1}$ is the recombination coefficient
(\nocite{n96}Nahar 1996) and $f_C = n_C/n_{C^+}$ is the neutral fraction. 
The ionisation rate, $\Gamma_{ion}$, is obtained by 
integrating the ionisation cross-section over the energy range 11.26 to 13.6 eV. 
For this integration, we used the radiation spectrum given by 
\nocite{d78}Draine (1978) and a constant ionisation cross section of 1.74 $\times 10^{-17}$ cm$^2$ 
(\nocite{np97}Nahar \& Pradhan 1997). 
If we assume that the spectrum of the background radiation is
independent of its integrated flux density i.e. G0, then Eq.~\ref{eqcion} can be used to estimate
$f_C$ for the G0 required for thermal balance.   The neutral fraction estimated
at a depth of Av/2 is 
0.08 for the model with $T_e = 60$ K and is 0.03 for the model with $T_e = 40$ K 
listed in Table~\ref{tab3}.  These are about
a factor of 10 higher than the neutral fractions inferred for CNM clouds (\lsim $3 \times 10^{-3}$;
\nocite{jt01}{Jenkins \& Tripp 2001) but not unreasonable for clouds
with \Htwo\ column density \NHtwo\ $\sim 10^{20}$ \cmtwo and G0 $\sim$ 5
(\nocite{htt91}Hollenbach \etal\ 1991).  

\subsection{Formation and Dissociation of molecular hydrogen in the R-C cloud}
\label{formdiss}

The properties of the R-C cloud discussed above can be used to examine the
formation and dissociation of \Htwo\ in the cloud. Conventionally, the rate
of formation of \Htwo\ is obtained from the frequency of collision
between H atoms and grains scaled by an efficiency factor
for recombination on the grain surface. The collision rates depend upon the temperature
and densities of H atoms and grains and the efficiency factor is 
estimated by making some reasonable assumptions regarding the properties
of the grains (\nocite{hs71}Hollenbach \& Salpeter 1971). The rate
of \Htwo\ formation per unit volume can be written as
(\nocite{vb86}van Dishoeck \& Black 1986) 
\be
R_{form}=3 \times 10^{-18}\; T_e^{0.5}\; n_H\; n_{HI}\; y_{ef} \mbox{ s}^{-1} \mbox{ cm}^{-3} 
\ee
where $y_{ef}$ is a parameter which takes into account the
sticking probability and formation efficiency. $y_{ef}$ is taken as unity for the
calculations presented here. The formation
rates obtained for the representative models vary from about $0.2\times10^{-11}$ s$^{-1}$\cmthree 
for the highest temperature model to about $12\times10^{-11}$ s$^{-1}$\cmthree for the
lowest temperature model.  These rates are given in Table~\ref{tab4}.
 
The \Htwo\ molecules in the cloud will be destroyed by FUV photons
(11 to 13.6 eV) and cosmic rays. Photodissociation is initiated
by line absorption (Lyman and Werner lines) and subsequent fluorescence
to the vibrational continuum of the ground state of \Htwo\ (P. M. Solomon
1965; private communication reported in \nocite{fsd66}Field, Somerville \& Dressler 1966,
\nocite{sw67}Stecher and Williams 1967).
Since the opacity to the UV lines from
the \Htwo\ molecule increases with depth (self-shielding) and several line transitions are 
involved, the dissociation rates at different depths are calculated numerically.  
Further, attenuation of FUV radiation field due to dust has to be taken
into account to calculate the dissociation rate.
An analytical approximation to the dissociation rate taking into account 
these effects is given by \nocite{db96}Draine \& Bertoldi (1996); 
\bea
R_{phdiss} = & (NH_2/10^{14})^{-0.75} e^{-4.0Av} \times \nonumber \\
             & 4.17 \times 10^{-11}\; G0\; n_{H2}\: \mbox{ s}^{-1} \mbox{ cm}^{-3}.  
\label{eqphdis}
\eea
Here $e^{-4.0Av}$ takes into account the dust 
attenuation near $\sim$ 12 eV. The self-shielding effect is absorbed in the term $(NH_2/10^{14})^{-0.75}$,
which is set to unity for \NHtwo\ $\le 10^{14}$ \cmtwo.  We
used the G0 estimated for thermal balance (see Table~\ref{tab4}) to determine 
the dissociation rate per unit volume at a depth of Av/2 in the cloud.
The estimated values for dissociation rate per unit volume (tabulated in
Table~\ref{tab4}) for the different models
are more than an order of magnitude smaller than the formation rate.
This difference in rates may indicate
that the R-C cloud is in the process of molecular formation similar to, for example,  
the HISA cloud G28.17+0.05 (\nocite{metal01}Minter \etal\ 2001).

We used the survey data of \nocite{detal87}Dame \etal\ (1987) to
investigate whether \tCO\ line emission is associated with the R-C cloud.
A \tCO\ line feature of similar LSR velocity and width as that of 
the observed \HI\ line toward the R-C cloud is present in some directions.
However, this \tCO\ line feature is not detected over the entire extent
of the R-C cloud. This may support the fact the molecular formation
in the R-C cloud is not complete. At the estimated rate of molecular formation 
in the R-C cloud, it should take \gsim $10^{5}$ years for converting all the 
\HI\ to \Htwo.


\section{Summary and Future Observations}
\label{sum}

In paper I, preliminary analysis of CRRL data obtained 
as part of a 327 MHz recombination line survey of the Galactic plane were presented. 
In this paper, we have for the first time, shown that the CRRL arising near
the Galactic centre within $l\sim10^{\circ}$ show excellent kinematic
correlation with the HISA features from the Riegel-Crutcher cloud arguing for
a common origin for the CRRL and HISA features.  The R-C cloud is a
HISA cloud located about 125 pc in the Galactic centre direction.  
Additionally, we have reported association of low frequency CRRL emission
with a few other HISA clouds in the inner Galaxy.  


We have also demonstrated that low frequency CRRL data at several frequencies
along with \HI\ observations can be used to constrain the physical properties of
the cold \HI\ regions. For the analysis presented here we made use of the CRRL observations
at 327 and 76 MHz along with \HI\ data to model the physical conditions in the
R-C cloud.   We find that models which fit
the 76 MHz and 327 MHz data and are constrained by the LOS size of the R-C cloud are the following:
$T_e$ $\sim$ 40 $\rightarrow$ 60 K, $n_e$ $\sim$
0.8 $\rightarrow$ 0.05 \cmthree and $S$ $\sim$ 0.03 $\rightarrow$ 3.5 pc.
The derived physical properties were used to examine 
the heating and cooling processes in the R-C cloud. The dominant heating and cooling processes were
found to be photoelectric emission and the \CII\ 158 \mum line emission respectively. 
The thermal balance between these two processes was used to constrain the 
diffuse FUV flux density on the cloud, which in Habing units (G0) ranges between $\sim$ 4 and 7. 
Further, we investigated the H$_2$ formation
and dissociation in the cloud and found that the formation rate per unit volume
exceeds the dissociation rate per unit volume by at least an order of magnitude. 
Based on this imbalance in the
formation and dissociation rate we conclude that the RC cloud is in the process
of converting from \HI\ to \Htwo\ and will convert all its atomic hydrogen into the
molecular form in a time scale \gsim $10^{5}$ years.  

The cold \HI\ gas observed as HISA features are ubiquitous
in the inner Galaxy and form an important part of the ISM. Our analysis 
shows that combining CRRL and \HI\ data can give important insight into the nature
of these cold gas. 

\begin{table*}
\small
\centering
\begin{minipage}{140mm}
\caption{Integration time for detecting CRRLs with SKA pathfinders \label{tab5}}
\begin{tabular}{lrrcccccc}
\hline
Telescope &  Freq   & $T_L$ & $\theta_r$ & N$_{base}$ & Tsys$^{1}$ & A$_{eff}$ & N$_{line}$ & $t_{int}$    \\ 
          &         &     &            &            &      &           &       &      \\
          & (MHz)   & (K) & (\arcmin)  &            & (K)  & (m$^2$)   &       & (hrs)     \\
\hline
MWA     &   95  & $-$6.1 & 22 & 30794 &  500 + 4200  & 23  & 30 & 22 \\ 
        &  200  & 1.1    & 10 & 30794 &  70  + 850  & 20  & 15 & 38 \\
MeerKAT$^{2}$ &  700  & 0.02   &  5 & 400$\times$ $f_c$   &  32  + 82  & 100 & 20 & 120/$f_c$  \\
ASKAP$^{3}$  &  750  & 0.02   &  5.6 & 49   &  50  + 58  & 90  & 10 & 1000 \\
\hline
\end{tabular}

\medskip
{$^1$ }{$T_{sys} = T_{rec} + T_{sky}$, where $T_{rec}$ is the receiver temperature and 
$T_{sky}$ is the contribution from the galactic background emission.} \\
{$^2$}{The MeerKAT antenna configuration details are taken from Booth \etal\ (2010). 
We have scaled $N_{base}$ approximately for the new 64 antenna configuration of MeerKat. Since
the scaling factor is not known well, we included a correction factor $f_c$ to indicate how 
the integration time changes with $f_c$.  The 20 recombination lines will 
span the frequency range $\sim$ 600 to 800 MHz.} \\
{$^3$}{The ASKAP parameters are taken from Gupta \etal (2008). The 10
recombination lines will span the frequency range $\sim$ 700 to 800 MHz.}
\end{minipage}
\end{table*}

We investigate the possibility of imaging the CRRL emission from \HI\ self-absorbing
clouds with upcoming Square Kilometre Array Pathfinders. The Murchison Widefield Array 
(MWA), Australian Square Kilometre Array Pathfinder (ASKAP) and the Karoo Array 
Telescope (MeerKAT) are considered for the investigation.
Observing CRRL emission with the Long Wavelength Array (LWA) is discussed by 
\nocite{petal10}Peters \etal (2010) and hence will not be discussed here. High angular resolution
observation with the upcoming arrays will help, for example, resolve the `line width problem'
(see Section~\ref{lwidth}). The integration times required to image CRRL emission
from the inner Galaxy with the different arrays are listed in Table~\ref{tab5}. The
carbon line temperature
is computed using the optical depth of $T_e = $ 50 K model given in Table~\ref{tab3}.
The peak line temperatures ($T_L$)  obtained from this model for 
line width of 12.2 \kms (see Table~\ref{tab2}) and galactic background temperature
of 500 K at 327 MHz are listed in Table~\ref{tab5}. This background temperature 
is an average value over
the angular resolutions of the interferometric observation. The RMS brightness
temperature ($\sigma_{RMS}$) in K for observations with the dual polarised interferometers is 
calculated using the equation
\be
\sigma_{RMS} = \frac{T_{sys}}{A_{eff} \sqrt{4 N_{base} N_{line} \Delta f\; t_{int}}} \frac{\lambda^2}{\theta_r^2}
\ee
where $T_{sys}$ is the system temperature in K, $A_{eff}$ is the effective area in $m^2$,
$N_{line}$ is the number of recombination lines that can be simultaneously observed,
$\Delta f$ is the frequency resolution in Hz corresponding to the line width of 12.2 \kms, $t_{int}$ is the
integration time in sec. $N_{base}$ is the number of baselines with length 
$\frac{\lambda}{\theta_r}$, where $\theta_r$ is the angular resolution of the image
and $\lambda$ is the observing wavelength. In the above equation the unit of
$\lambda$ is meter. Integration times listed in Table~\ref{tab5} are for detecting
CRRLs at 4$\sigma$ level. The estimated values show that imaging 
CRRLs  with the MWA is feasible.

\section{Acknowledgements}

Late Prof. K. R. Anantharamaiah first suggested that we conduct the recombination 
line survey near 327 MHz with the Ooty Radio Telescope. We are grateful to Anantha for
his guidance and support, to which we owe much of the work we have
accomplished in our careers. DAR thanks F. J. Lockman and D. S. Balser,
NRAO, Green Bank, for the many stimulating discussions and helpful
suggestions. DAR is thankful to NCRA
for providing local hospitality while a part of the work on the paper was done.
DAR also thanks Resmi Verma, a visiting student at RRI, for helping with 
obtaining the 2\ddeg\ averaged \HI\ spectra and Divya Oberoi, MIT, Haystack
Observatory for providing the MWA configuration details. We thank the 
referee Naomi McClure-Griffiths for comments which have helped improve
the paper and for suggesting to include feasibility of observing CRRLs
with the SKA Pathfinders.


\end{document}